# Pulsed loading of a magneto-optical trap on atom chip for fast pressure recovery in ultrahigh vacuum environment

Vivek Singh*,[1,2,a)] V. B. Tiwari,[1,2] A. Chaudhary,[1] S. Sarkar,[1,2] and S. R. Mishra[1,2]

[1)]*Laser Physics Applications Division, Raja Ramanna Centre for Advanced Technology, Indore-452013, India*

[2)]*Homi Bhabha National Institute, Anushaktinagar, Mumbai-400094, India*

(Dated: 18 March 2024)

This study presents investigations on pulsed loading of a magneto-optical trap (MOT) on an atom chip in an UHV environment. Using three parallel resistively heated Rb-metal dispensers activated by pulsed current supply, approximately $3.0 \times 10^7$ cold $^{87}Rb$ atoms were loaded into the MOT. A current pulse of $\sim 24$ A with duration of $\sim 10$ s raised the pressure in the chamber from $2.0 \times 10^{-10}$ Torr to $3.3 \times 10^{-10}$ Torr. Remarkably, the pressure recovery time after switching off the dispensers current was found to be $\sim 600$ ms, making a significant advancement in achieving fast recovery of UHV environment surrounding the MOT region. This study is very useful for laser cooling and magnetic trapping / evaporative cooling of atoms on atom chip in the same UHV chamber.

---

[a)]Electronic mail: viveksingh@rrcat.gov.in

## I. INTRODUCTION

In contemporary times, the magneto-optical trap (MOT) has emerged as a versatile tool for producing cold atom samples with the temperature range spanning from few tens to few hundreds of micro-Kelvins. These cold atoms, originating from the MOT, serve as the basic entity for various fundamental studies including Bose-Einstein condensation, cold collisions, optical lattices, precision spectroscopy, etc. The cold atoms from MOT are also used for making atom-optic devices such such as atomic clocks[1,2], atomic gravimeters[3–5], atom-gyroscopes[6], and atomic qubits for quantum information[7,8].

For experiments like magnetic or optical dipole trapping of cold atoms from MOT, one needs an ultrahigh vacuum (UHV) environment in the space surrounding the MOT. But loading a MOT from the background vapor requires a higher vapor pressure ($\sim 10^{-8}$ Torr) in the chamber, which is conflicting to the UHV requirement for magnetic or dipole trap lifetime. A solution to this problem is to load MOT in UHV chamber by Zeeman slower technique[9] or use a double-MOT setup[10] where atoms in a vapor chamber MOT are transferred to UHV chamber MOT for further trapping and study. But both these methods increase the complexity of the experimental system. If MOT is loaded from background vapor at UHV pressure in single chamber setup, the MOT loading time is very long (25-65 second)[11,12]. This extended loading duration adversely affects the experiment's duty cycle, which is not desirable in several cases. Therefore, loading of a MOT in UHV environment by operating Rb-vapor source in pulsed mode is an attractive alternative, if the UHV environment in the chamber, which gets degraded due to emission of gases from dispensers, can be quickly recovered after the MOT loading. Several researchers have tried this type of pulsed loading of MOT in UHV environment[13–16]. However, the timescale for the recovery of the original UHV pressure after MOT loading ranges from 3 to 4 seconds[17–22]. With a custom-designed heat sink, this recovery time for the UHV in the chamber was reduced to approximately 100 ms for a pulsed loading of MOT[13].

Here, we report our studies on pulsed loading of a MOT on atom chip to achieve a fast recovery of UHV environment in the chamber after the MOT loading. We have used three resistively heated Rb-metal dispensers connected in parallel configuration which is helpful for faster heat dissipation. A current pulse of $\sim 24$ A with duration of $\sim 10$ s is applied to combined dispensers assembly such that each Rb dispenser operates at nearly 8 A of current. With the application of current pulse, the pressure in the chamber rises from $2.0 \times 10^{-10}$ Torr to $3.3 \times 10^{-10}$ Torr due to emission of Rb vapor and other gases from the dispensers. Nearly $3.0 \times 10^7$ cold $^{87}Rb$ atoms were loaded in the MOT during this pulsed operation. The UHV pressure recovery time in the chamber, after switching-off the dispensers current, was approximately 600 ms. Such a low recovery time for UHV after the MOT loading has not been reported earlier to the best of our knowledge in similar dispensers configuration without using any heat sink. Therefore, this approach of MOT loading is very useful for further experiments like magnetic trapping and evaporative cooling of atoms on atom chip. With our approach, the MOT is loaded in UHV chamber without complicating the design of the setup. Also, the UHV background pressure is quickly recovered after the MOT loading.

## II. EXPERIMENTAL SETUP

The vacuum chamber was evacuated by using several vacuum pumps, including a turbo molecular pump (TMP), a titanium sublimation pump (TSP), and a sputter ion pump (SIP). The final base pressure in the chamber was $\sim 2.0 \times 10^{-10}$ Torr. This low pressure was achieved following a five-day baking process of the vacuum system. Further details of the experimental arrangement can be found in other references[11,12]. A two-pin vacuum feedthrough (DN 35 CF) having three Rb dispensers (Rb/NF/3.4/12FT) welded in parallel on copper rods of feedthrough, was utilized as shown in Fig. 1 (a). In this arrangement, we have not made any special provision for heat sink as compared to other reported work[13]. This feedthrough was inserted inside the vacuum chamber through a viewport hole, positioning the dispensers approximately 17 cm away from the center of the octagonal science chamber. The generation of Rubidium vapor in the vacuum chamber was achieved by passing a dc current through this two-pin feedthrough.

Fig. 1(b) shows the schematic of the experimental setup for U-MOT formation near the atom chip surface. Due to reflection on the chip surface, the two MOT beams oriented at $45°$ (illustrated in Fig. 1(b)) give rise to four MOT beams within the overlapping region. Additionally, two counter-propagating MOT beams (not depicted in Fig. 1(b)) in the x-direction (perpendicular to plane of paper) were employed to achieve the necessary six MOT beams in the overlapping region. To generate the required quadrupole-like magnetic field pattern for MOT formation, a U-shaped thick copper wire with a rectangular cross-section was positioned behind the chip (as depicted in Fig. 1(b))[11]. The magnetic field due to current in this U-wire, combined with the magnetic field due to the bias coils, resulted in the desired quadrupole field for MOT formation.

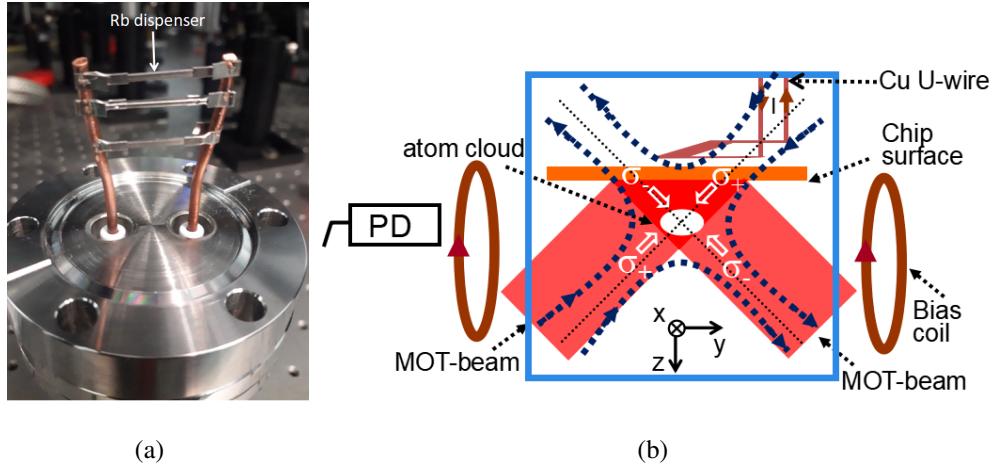

FIG. 1. (a) The photograph of the assembly of three Rb dispensers connected in parallel configuration using the spot welding on copper rods of vacuum feedthrough. (b) The schematic of the experimental setup for U-MOT formation near the chip surface. The dotted curves show the generated quadrupole like magnetic field lines due to current in U-shaped copper wire and bias magnetic fields. The MOT is formed $\sim$ 6 mm below the atom-chip surface.

An FPGA-based electronic controller is employed to generate different analog, isolated analog, and digital pulses, enabling the control of various equipment such as acousto-optical modulators, power supplies, CCD cameras, mechanical shutters, photo-diodes etc. This controller facilitated the execution of different experimental sequences, ranging from the current pulse in Rb dispenser to loading of the MOT. Four independent MOT laser beams, combined with re-pumper beams, are utilized for operation of this MOT, which is called U-MOT due to use of current in U-wire for MOT magnetic field. The intensity of each cooling beam is approximately 15 $mW/cm^2$, and the laser beam detuning is set at -14 MHz for U-MOT formation. When operating the MOT in proximity to the atom-chip surface, a quadrupole field is generated by passing a current of 60 A through a U-shaped copper wire, in conjunction with the bias fields ($B_y \sim$ 10 G and $B_z \sim$ 1.5 G).

In the experiments, different current pulses have been tried on Rb dispensers assembly to know the most suitable shape of current pulse for MOT loading and fast UHV recovery in the chamber. For this, the rise and fall of temperature of dispensers was also recorded after application of current pulse. The Fig. 2 shows the temporal variation of the temperature during the heating of dispenser due to current pulse. The temperature of the dispenser source was derived by observing the thermal power radiated by the dispensers on photodiode. The thermal radiation power (S) observed on a photodiode depends[18] on the temperature (T) as S $\propto T^4$. Fig. 2 (a) shows the rise in temperature at lower current with longer time duration. As shown in Fig. 2(a), for lower current, the temperature increases slowly until dispenser reaches the thermal equilibrium. After the switching-off the dispenser current, the temperature decreases to its initial value within a few seconds. In contrast to this, with higher current and short duration pulse, as shown in Fig. 2 (b), the rise in dispenser temperature is faster. However, after switching-off the current, the initial decay in the temperature is faster followed by a similar slower decay of the temperature as for the case of low current pulse. This fast reduction in dispenser temperature (in Figs 2(a) and (b)) after the end of current pulse is an important and useful feature in our experiments for the pulsed operation of Rb-dispenser.

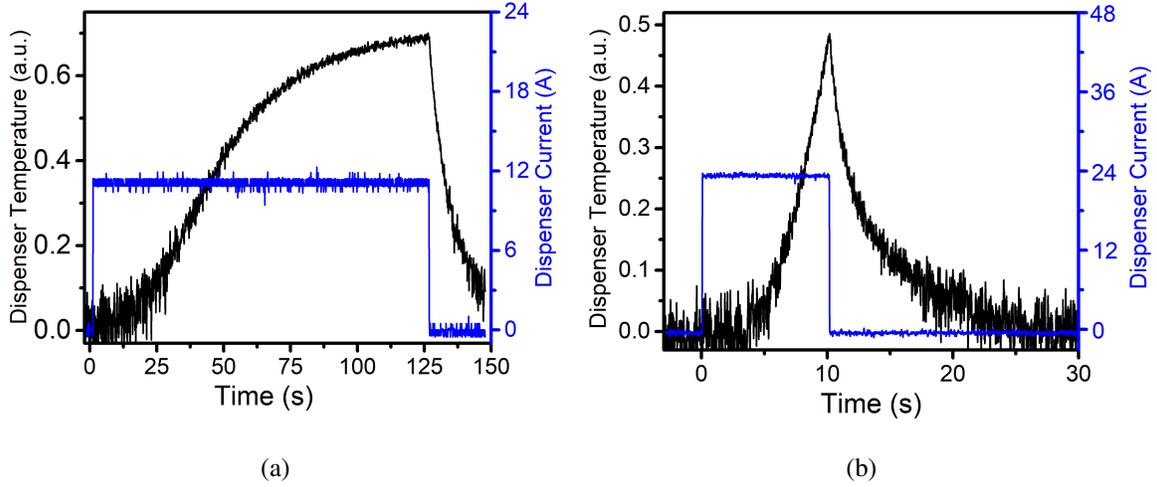

(a)  (b)

FIG. 2. The change in dispenser temperature with time, as derived from thermal radiation signal from dispenser, for different DC current pulses (blue color curves) applied to dispensers assembly. (a) The dispenser current pulse of lower amplitude (11 A) and longer duration (125 s). (b) The dispenser current pulse of higher amplitude (24 A) and shorter duration (10 s).

Since Rb metal dispenser has a threshold behaviour for emission of the atoms, the emitted Rb flux gets stopped after a temperature and UHV recovery process can start from there. For sensing the emission of rubidium atoms from the dispenser after application of the current pulse, the fluorescence from the $^{87}Rb$ atoms in the MOT chamber was monitored on a photodiode in a synchronous manner to the current pulse. The Fig. 3 shows the fluorescence signal from the atoms in the chamber as function of time for different shapes of the current pulse. Here, the photodiode fluorescence signal is proportional to rise in atom number density (or Rb vapour pressure) due to emission of Rb atoms from dispenser. Fig. 3 (a) shows the fluorescence recorded for a composite current pulse 24 A, 10 s and 12 A, 4 s. It is observed that very low atomic flux is emitted during the initial ∼ 8 seconds. After this time, the Rb dispenser gets heated adequately above the threshold to start the emission of Rb atoms from it. After this delay, fluorescence ( or number of atoms) rises quickly as shown in Fig. 3 (a). At the end of the current pulse, the fluorescence from Rb atoms decays exponentially after the switching-off the current in the dispenser. The exponential decay time is $\sim$ 1.2 s for Fig. 3(a). Fig. 3(b) shows the fluorescence for 24 A, 10 s current pulse. The decay time constant for this current pulse is $\sim$ 600 ms after the switching-off the current in the dispensers. It is evident that this current pulse gives faster recovery of the UHV pressure in the chamber as compared to earlier (Fig. 3(a)) pulse duration. For both the current pulse configuration, the vacuum pressure is increased from $2.0 \times 10^{-10}$ Torr to $3.3 \times 10^{-10}$ Torr as read by SIP controller.

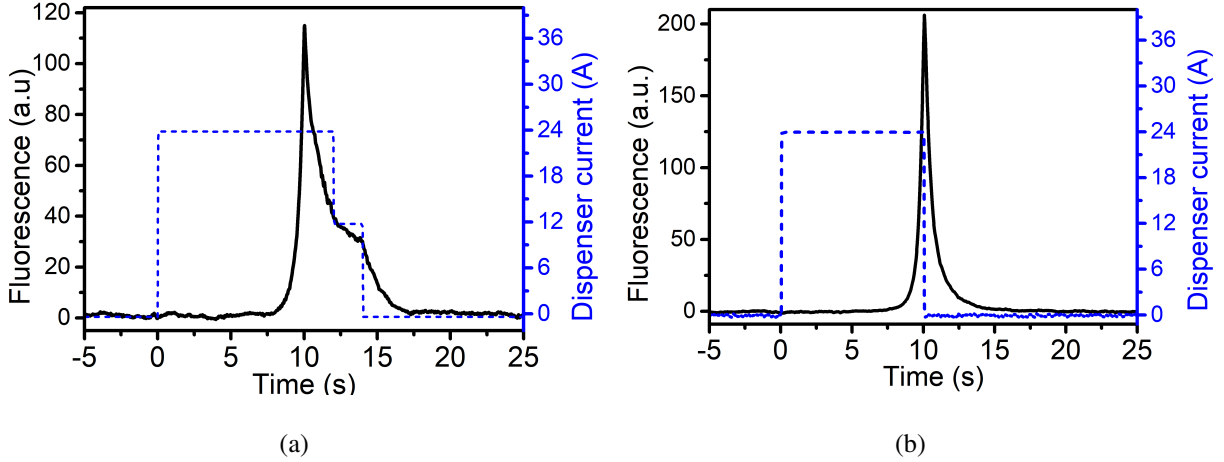

(a)  (b)

FIG. 3. (a) The measured fluorescence signal (black curve) from the rubidium atoms in the chamber for a composite current pulse (24 A, 10 s and 12 A, 4 s, shown by blue dashed curve). (b) The measured fluorescence signal (black curve) from the rubidium atoms in the chamber for a shorter current pulse (24 A, 10 s, blue dashed curve). The fluorescence decay times (1/e) after the current pulse are $\sim$ 1200 ms and $\sim$ 600 ms for cases (a) and (b) respectively.

We have demonstrated the U-MOT loading using various current pulses for the dispensers. Fig. 4(a) shows the loading of U-MOT graph for current pulse of 24 A, 10 s. For the full loading of MOT, $\sim$ 2 seconds time is required for this current pulse. Nearly $3 \times 10^7$ $^{87}Rb$ atoms are trapped in the U-MOT. As shown in Fig. 4(b), the lifetime of atoms in MOT ($\sim$ 58 s) is very long indicating the good UHV in the chamber background.

The Table I shows the results of MOT loading and UHV pressure recovery time for different current pulses applied to dispensers assembly in our setup. We note that, for the same peak current, the longer pulse duration gives higher pressure rise (i.e. degradation of UHV) in the chamber but more number of atoms loaded in the MOT. Table I also represents that UHV pressure recovery time gets further reduced for smaller current pulse duration with lesser number of trapped atoms in MOT. By increasing the current and making the pulse shorter might lead to faster pressure recovery but it may also damage the dispenser due to high temperature rise in dispenser. The use of parallel configuration of dispensers, as we used in our experiments, is more effective than series configuration because of better thermal heat dissipation achievable in the parallel configuration.

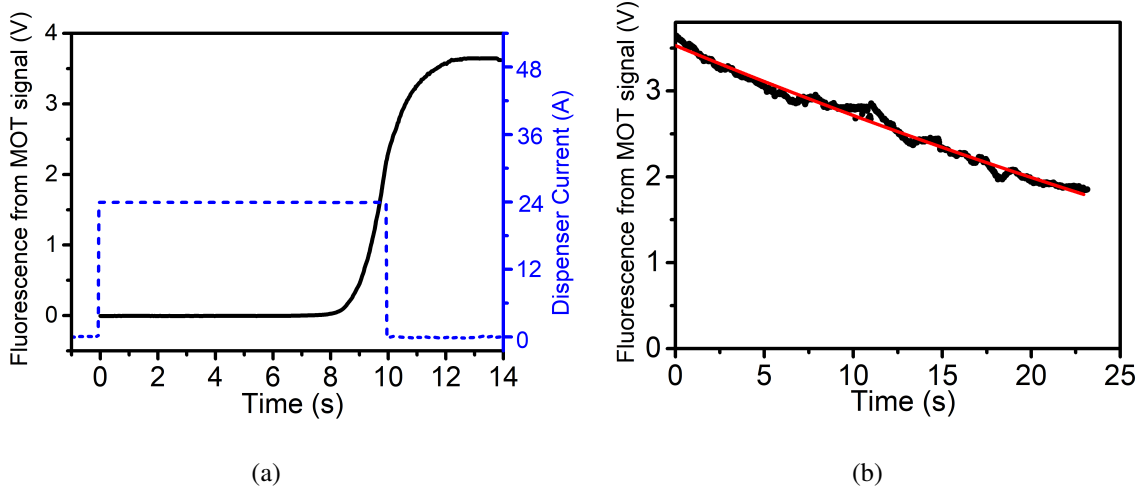

(a) (b)

FIG. 4. (a) The MOT fluorescence signal during MOT loading with a pulsed operation of dispenser source. The DC current pulse (24 A, 10 s) applied to dispensers is indicated by dashed curve in figure (a). (b) Decay of fluorescence from MOT atoms indicating the MOT life-time $\sim 58$ s. The red curve in figure (b) is exponential fit to experimental data.

TABLE I. Table summarizing the UHV pressure recovery time, number of atoms in MOT and elevated UHV pressure for different current pulses applied to dispensers assembly for the MOT loading.

| Current pulse parameters | Elevated pressure | Number of atoms in MOT | UHV Pressure recovery time |
| --- | --- | --- | --- |
| 24 A, 9 s | $2.3 \times 10^{-10}$ Torr | $8.0 \times 10^6$ | 485 ms |
| 24 A, 9.5 s | $2.6 \times 10^{-10}$ Torr | $2.0 \times 10^7$ | 548 ms |
| 24 A, 10 s | $3.3 \times 10^{-10}$ Torr | $3.0 \times 10^7$ | 600 ms |
| 24 A,10 s,12 A, 4 s | $3.3 \times 10^{-10}$ Torr | $5.5 \times 10^7$ | 1200 ms |

## III. CONCLUSION

The pulsed loading of a magneto-optical trap (MOT) on an atom chip in UHV environment has been demonstrated with a fast recovery of UHV pressure after the MOT loading. This is achieved by using three Rb dispensers in parallel configuration driven by the DC current pulse. The experimental results show that a fast decrease in dispenser temperature is essential for fast recovery of UHV in the chamber. Approximately $3.0 \times 10^7$ cold $^{87}Rb$ atoms were loaded into the MOT with a fast (in 600 ms) recovery of UHV in the chamber by using the dispensers in parallel configuration. This pulsed loading of MOT is useful for performing further experiments such as magnetic trapping and evaporative cooling in UHV environment.

## IV. ACKNOWLEDGEMENT

We acknowledge the help provided by S. Supakar during the experiments. We are also thankful to S. P. Ram for his helpful technical suggestions.